\documentclass[11pt,a4paper]{JHEP3}
\usepackage{amsmath, amssymb}
\usepackage{euscript}
\def\be{\begin{eqnarray}}
\def\ee{\end{eqnarray}}
\def\0{\nonumber}

\usepackage{amsfonts}
\usepackage{verbatim}
\usepackage{epsfig}
\usepackage{graphicx}

\def\0{\nonumber}

\def\a{{\bf a}}
\def\b{{\bf b}}
\def\c{{\bf c}}

\preprint{SISSA/34/2010/EP\\\tt hep-th/xxxx.yyyy}

\title{Non--simply laced Lie algebras via F theory strings}

\author{L. Bonora, R. Savelli \\
International School for Advanced Studies (SISSA/ISAS)\\
Via Bonomea 265, 34136 Trieste, Italy, and INFN, Sezione di
Trieste\\
 
E-mail:   \email{bonora@sissa.it}, \email{savelli@sissa.it}}

\abstract{In order to describe the appearance in F theory of the non--simply--laced Lie algebras, we use the representation of symmetry enhancements by means of string junctions. After an introduction
to the techniques used to describe symmetry enhancement, that is algebraic geometry, BPS states analysis and string junctions, we concentrate on the latter. We give an explicit description
of the folding of ${\bf D_{2n}}$ to ${\bf B_n}$, of the folding of ${\bf E_6}$ to ${\bf F_4}$ and that of ${\bf D_4}$ to ${\bf G_2}$ in terms of junctions and Jordan strings. We also discuss the case of ${\bf C_n}$, but we are unable in this case to provide a string interpretation.}

\keywords{F theory, (p,q) strings, roots}

\begin{document}

\maketitle

\section{Introduction}

The recent return of interest in F theory, \cite{vafa}, has originated from the suggestion of possible low energy phenomenological implications, more precisely by the possibility to accommodate in this theory a gravity decoupling at the scale of grand-unification together with low energy effective grand-unified actions which extend the MSSM, \cite{ftheory}. While the coexistence of these conditions is still under scrutiny, we are interested here in one of the characteristics of F-theory that makes the previous conjectures plausible. We refer to the symmetry enhancements that can occur in it, which allow for virtually all types of gauge symmetries, that is all type of gauge Lie algebras (with possible bounds only on their rank). What is most interesting for the above mentioned phenomenological applications is in particular the possibility to accommodate theories characterized by the series of exceptional simply--laced Lie algebras. But in fact all Lie algebras can be realized, also the non--simple--laced ones. 

Independently of its possible phenomenological applications, F theory vacua are characterized by peculiar aspects that distinguish them from other superstring vacua. Generally speaking F theory vacua are more `constrained' than others. In particular the number of 7--branes, their type and, eventually, the type of enhanced symmetry is a result of the dynamics (geometry) rather than put in by 
hand, as is the case of other compactifications with branes. The price for it is that, generically, the relevant open strings are mutually non-perturbative. This  is not to say, however, that nothing can be said about, for instance, the dynamics in 4D, as refs.\cite{ftheory} abundantly testify. It is therefore important to analyze and understand the dynamics of F-theory. A lot has already been done in the past, but there are still aspects of the theory where the analysis has not been completed. To mention one important problem, we do not know 
what form the Freed--Witten anomaly takes in such a non-perturbative context. Even though our research originated from this problem,
in this paper our aim will be more modest: we will concentrate on symmetry enhancement in F theory and, in particular, on the appearance of non--simply--laced Lie algebras. 

The symmetry enhancement in F theory can be analyzed with various (complementary) techniques: either with algebraic geometric techniques (Tate's algorithm) \cite{tate, BIKMSV},  or by studying the BPS strings stretched among 7--branes \cite{joha, zwie1, zwie2, BH}, or by means of the (strictly related) Lie algebra realization via string junctions \cite{zwie3}. Below we will focus on the last method. Our purpose in this paper is to apply it to the analysis of non--simply--laced Lie algebras. We will show in particular how to obtain a description of the root system of the latter by means of F theory string junctions, attached to a system of (in general, mutually non-perturbative) 7-branes.  

The paper is organized as follows. We will start in the following section with a short review of symmetry enhancement in F theory, and continue in section 3 with a review of the string junction technology needed in the sequel. Then we will begin with the analysis of the folding of different Lie algebras: in section 4 we will consider the folding of ${\bf D_{2n}}$ to ${\bf B_n}$, in section 5 the folding of ${\bf E_6}$ to ${\bf F_4}$, in section 6 that of ${\bf D_4}$ to ${\bf G_2}$. Finally in section 7 we will 
sketch the method to obtain ${\bf C_n}$. Our first purpose throughout is to show that such foldings can be formulated in terms of junctions, stretching among a given set of (in general) mutually non-perturbative 7-branes. The next crucial aim is to provide an interpretation of the physical states in terms of Jordan strings (by undoing the corresponding junctions) so as to render self--evident the symmetry of the Dynkin diagram responsible for the foldings. While we are  
able to do this for all the cases considered above, this turned out to be impossible for
the ${\bf C_n}$, for which we were not able to provide a string interpretation.

\section{A concise review of symmetry enhancement in F theory}

\subsection{Geometric perspective}\label{alggeo}

The purpose of this section is a flash review of symmetry enhancement in two typical cases of F theory compactifications, focusing in particular on the algebraic mechanism responsible for the appearance of non--simply--laced gauge groups as opposed to simply--laced ones.
We will consider two possible geometrical schemes: either F theory compactified on $R_{1,7} \times \textrm{K3}$, where the K3 is elliptically fibered over a 2-sphere, or F theory compactified down to 4D on a CY fourfold $X_4$.  $X_4$ is elliptically fibered over a complex threefold $B_3$ and it admits the (standard) Weierstrass representation, \cite{tate, BIKMSV}
\be
y^2 +a_1 xy + a_3 y= x^3 + a_2 x^2 +a_4 x +a_6 \label{weiestrass}
\ee 
where $x,y$ are affine coordinates parametrizing the elliptic fiber and $a_1,a_2,a_3,a_4,a_6$ are locally defined polynomials.
In the first scheme mentioned above the latter depend on a unique coordinate $z$ which spans the Riemann sphere. In the second they will depend on several local coordinates. In this case the appropriate method to study symmetry enhancement is Tate's algorithm. The elliptic fibers degenerate over specific divisors of the base $B_3$. Let $\Sigma$ be one such divisor and let $\sigma=0$  be its local defining equation. If the discriminant $\Delta$ of the curve (\ref{weiestrass}) vanishes on it, the corresponding fiber degenerates; if, for instance, $\Delta$ is divisible by $\sigma$ and not by $\sigma^2$, we have at $\sigma=0$ a singularity of Kodaira type $I_1$, which does not constitute a singularity of the total space of the fibration and does not give rise to non-abelian enhanced symmetry. In order to come across the latter we need more severe singularities. The virtue of Tate's algorithm is that it enables us to classify all possible singularities of 
(\ref{weiestrass}) in a systematic tree--like way by analyzing the increasing order of zeros of $\Delta$ and of suitable polynomial combinations of $a_1,a_2,a_3,a_4,a_6$.  The first non trivial instance is when the polynomials $a_1,a_2,a_3,a_4,a_6$  are such that  (\ref{weiestrass}) takes, at leading order near the origin, the form of the following \emph{quadratic} equation in $x,y,\sigma$:
\be\label{quad}
y^2+a_1xy+a_{3,1}\sigma y=a_2x^2+a_{4,1}\sigma x+a_{6,2}\sigma^2
\ee
and $\Delta$ has a double zero at $\sigma=0$. In the case the quadratic form of (\ref{quad}) is non-singular, one blows up the origin and resolves the singularity there. The exceptional divisor of such a resolution is given by the following generically irreducible non-singular quadratic equation:
\be\label{excep}
y_1^2+a_1x_1y_1+a_{3,1}y_1=a_2x_1^2+a_{4,1} x_1+a_{6,2}
\ee
where $x_1=x/\sigma$ and $y_1=y/\sigma$ parametrize in a patch the lines through the singular point and as such, together with $\sigma$, they are local coordinates in the neighborhood of the exceptional divisor, which is placed at $\sigma=0$. This is a singularity of Kodaira type $I_2$, which gives rise to an enhanced symmetry of the $SU(2)$ type.

Let us continue along the same branch of the algorithm for simplicity: at the next step, one encounters two possibilities that differentiate between the non--simply laced and the simply laced alternative for the gauge symmetry.

\begin{itemize}

\item Require $a_{3,1}, a_{4,1}, a_{6,2}$ to be further divisible by $\sigma$, so that (\ref{quad}) becomes:
\be\label{quad1}
y^2+a_1xy+a_{3,2}\sigma^2 y=a_2x^2+a_{4,2}\sigma^2 x+a_{6,3}\sigma^3
\ee
Now, by performing the same blow up as before, we end up with \emph{two} exceptional divisors represented on each point of the base by the two lines solving the equation
\be
y_1^2+a_1x_1y_1-a_2x_1^2=0\label{quadratic}
\ee
They are not globally defined in general and they will experience monodromy as one goes along a closed path on the base. This singularity type is named $I_3^{ns}$ ($ns$ standing for non-split) and gives rise to unconventional gauge symmetry.\\
One can go further along this sub-branch of the algorithm just requiring divisibility by $\sigma$ of $a_{6,3}$: this  induces an $SU(2)$ singularity at the origin $(x_1=y_1=\sigma=0)$ that survives after the blow up, so that a second blow up is necessary to completely resolve the singularity, which leads to a further irreducible exceptional divisor, similar to (\ref{excep}),  placed at $\sigma=0$ in the coordinate chart $(x_2=x_1/\sigma, y_2=y_1/\sigma, \sigma)$ This is the type $I_4^{ns}$ singularity, which corresponds to the $Sp(2)$ gauge group.\\
By induction, one easily constructs in this way the series of  ${\bf C_n}$ algebras, the resolution of the corresponding singularities being characterized by $n-1$ pairs of non-split exceptional divisors plus an irreducible one; moreover one has also a tower of unconventional gauge symmetries for which there are instead $n$ pairs of non-split exceptional divisors. 

\item Require $a_2, a_{4,1}, a_{6,2}$ to be further divisible by $\sigma$, so that (\ref{quad}) becomes:
\be\label{quad2}
y^2+a_1xy+a_{3,1}\sigma y=a_{2,1}\sigma x^2+a_{4,2}\sigma^2 x+a_{6,3}\sigma^3
\ee
Blowing up in this case leads to a resolution of the singularity by means of two globally distinct (split) exceptional divisors, described by the equation:
\be
y_1(y_1+a_1x_1+a_{3,1})=0\label{quadratic1}
\ee
This singularity type is named $I_3^{s}$ ($s$ standing for split) and gives rise to the familiar SU(3) gauge symmetry.\\
Again one can go on with the algorithm just requiring $\sigma$ to divide $a_{3,1}$ and $a_{6,3}$. Again this will induce a residual $SU(2)$ singularity at the origin, i.e. the intersection of the two previously found split exceptional divisors: the blow up of such singularity will lead as before to an additional irreducible exceptional divisor, like (\ref{excep}), placed at $\sigma=0$ in the coordinate chart $(x_2, y_2, \sigma)$. This is the type $I_4^{s}$ singularity, which corresponds to the $SU(4)$ gauge group.\\
By induction, one constructs this way the entire series of ${\bf A_n}$ algebras, producing, out of the resolving procedure, $n/2$ pairs of split exceptional divisors if $n$ is even and $(n-1)/2$ pairs of split exceptional divisors plus an irreducible one if $n$ is odd. 

\end{itemize}

While the full classification of symmetry enhancements can be found in \cite{BIKMSV}, let us write here the polynomials whose factorization distinguishes the split case from the non-split one, also in the other branches of the algorithm, which are relevant for us because they contain the orthogonal and the exceptional gauge symmetries. 

\begin{enumerate}

\item  For the Kodaira singularity of type $IV^{\star ns}$ (corresponding to the gauge group $F_4$), the relevant polynomial is:
\be
y_2^2+a_{3,2}y_2-a_{6,4}=0\label{effe4}
\ee
which factorizes globally if we just require $a_{6,4}=0\,\textrm{mod}\, \sigma$, thus generating the $IV^{\star s}$ singularity (namely the $E_6$ gauge group).

\item For the Kodaira singularities of type $I^{\star ns}_{2k-3}$, $k\ge2$ (corresponding to the gauge groups $SO(4k+1)$), the relevant polynomials are:
\be
y_k^2+a_{3,k}y_k-a_{6,2k}=0\label{so1}
\ee
which factorize globally if we just require $a_{6,2k}=0\,\textrm{mod}\, \sigma$, thus generating the $I^{\star s}_{2k-3}$ singularities (namely $SO(4k+2)$ gauge groups).

\item For the Kodaira singularities of type $I^{\star ns}_{2k-2}$ (corresponding to the gauge groups $SO(4k+3)$, $k\ge2$), the relevant polynomials are:
\be
a_{2,1}x_k^2+a_{4,k+1}x_k+a_{6,2k+1}=0\label{so3}
\ee
for which we cannot change coordinates in order to formulate their factorization as before in terms of the vanishing mod $\sigma$ of some polynomial; anyway, if they factor, the associated singularities become $I^{\star s}_{2k-2}$, $k\ge2$ (namely $SO(4k+4)$ gauge groups).

\item Finally, the Kodaira singularity $I^{\star ns}_0$ (corresponding to the gauge group $G_2$) contains a subtlety. The relevant polynomial is:
\be
x_1^3+a_{2,1}x_1^2+a_{4,2}x_1+a_{6,3}=0\label{so8}
\ee
which describes a triple of non--split exceptional divisors. Clearly (\ref{so8}) can either partially or completely split. The former situation is achieved simply requiring  $a_{6,3}=0\,\textrm{mod}\, \sigma$, which leads to the so called type $I^{\star ss}_0$ ($ss$ standing for semi-split), corresponding to $SO(7)$ gauge group (a couple of non--split exceptional divisors and a split one). The latter is obtained by further requiring the factorization mod $\sigma$ of $x_1^2+a_{2,1}x_1+a_{4,2}$, which leads to three split exceptional divisors, but, as at point 3., cannot be formulated in terms of the vanishing mod $\sigma$ of some polynomial: this is the case of type $I^{\star s}_0$ (namely $SO(8)$ gauge group).

\end{enumerate}

In the above algebraic geometric description, what establishes the connection between the specific singularity and the enhanced symmetry is the fact that the intersection matrix of the components of each singular fiber (namely the various exceptional divisors we have found) is observed to take the form of the affine Cartan matrix of the corresponding non--Abelian Lie algebra. As we saw, this gives rise to the {\bf ADE} series of Lie algebras provided that no monodromy acts on the collapsing cycles\footnote{Actually if monodromies are present but they are all given by elements of the Weyl group (inner automorphisms), they can be undone by a gauge transformation in the fiber and thus we don't break the initial simply--laced gauge group.}. On the other hand when the opposite occurs and, in particular, when, going around the singularity, we pick up an outer automorphism of the Lie algebra, the gauge group gets orbifolded as we shrink the 2-cycles of the resolution to zero-size, and one ends up with a reduced gauge symmetry. These reductions via outer automorphisms are known, in Lie algebra theory, to be connected to the symmetry of the relevant Dynkin diagrams and to lead to the non--simply--laced algebras; precisely, a ${\mathbb Z}_2$ orbifold leads from ${\bf A_{2n-1}}$ to ${\bf C_n}$, from ${\bf D_n}$ to ${\bf B_{n-1}}$ and from ${\bf E_6}$ to ${\bf F_4}$, while the triality of the ${\bf D_4}$ Dynkin diagram leads to ${\bf G_2}$. 

\subsection{String perspective}

In the framework of algebraic geometry, the above is as much as one can say about the connection between singularity theory and 
enhancing of gauge symmetry (although some attempts were made in the past to render it more explicit, \cite{BRZ,Belhaj}). The gauge interpretation 
is supported by the duality with heterotic theory, when the latter exists. But, needless to say, a more direct and physical interpretation is clearly desirable and was indeed put forward in the early stage of F theory. It was based on the analysis of BPS spectrum of 7-branes. The spectrum of $[p,q]$-7--branes is formed by 
$\left(\begin{matrix}p\\ q\end{matrix}\right)$-strings. Since, in general, enhanced symmetry requires an 
assemblage of branes with different $p,q$ charges, it is evident that the strings that enter the game will in general be mutually non-perturbative. The search for 
BPS string states was carried out in refs. \cite{joha,zwie1,zwie2, BH} in the first scheme
referred to above, that is when F theory is compactified on $R_{1,7} \times \textrm{K3}$. 

In this case many things simplify. The Weierstrass representation can be written
in the traditional form 
\be
y^2= x^3+f(z) x +g(z)\label{weies2}
\ee
where $z$ is the coordinate on the ${\mathbb P}^1$ base of the elliptically fibered K3, and $f$ and $g$ are polynomials of degree eight and twelve, respectively. As usual the zeroes of the discriminant $\Delta = 4f^3+ 27 g^2$ identify the locations of 7--branes on the sphere ${\mathbb P}^1$. This compactification scheme gives rise, via collapsing of 7-branes, to many examples of symmetry enhancements, but it cannot give rise to non--simply--laced gauge groups, due to the absence of non-trivial monodromies on $\mathbb{R}^{1,7}$. The complex axi--dilaton field $\tau$ is implicitly defined by the equation
\be
j(\tau(z)) =4(24 f)^3/\Delta\0
\ee
where $j$ is the standard modular function, which maps the fundamental region of $\tau$, with respect to the $SL(2,Z)$ action, to the sphere. Once $\tau$ and the sites of 7--branes on the sphere are known, one can write down the metric 
on ${\mathbb P}^1$ \cite{SVY}:
\be
ds^2= Im(\tau) |\eta(\tau)|^4 \prod_i |z-z_i|^{- 1/6} dz d\bar z\label{metricds2}
\ee
where $\eta$ is the Dedekind function. Knowing the metric, one can, at least in principle, compute the geodesics.  
But one has to take into account also the string tension, which, for a 
$\left(\begin{matrix}p\\ q\end{matrix}\right)$ string, is 
\be
T_{p,q} = \frac 1{Im(\tau)} |p-q \tau| \label{tension}
\ee
It is therefore natural to define the effective length $ds_{p,q} = T_{p,q} ds $.
It measures the mass of the stretched strings between different branes.
The idea is to consider an allowed configuration of $[p,q]$ branes, which will 
eventually collapse, and produce the desired enhanced symmetry, and analyze all possible strings stretched between them. These strings, before collapse, will be massive. The states with minimal $ds_{p,q}$ length are recognized as BPS states
and will identify the massless gauge fields. 

The determination of all BPS string states is in principle possible; in practice it is not easy. One reason is that $\tau(z)$ is, in general, a function defined only implicitly, so that only numerical techniques are viable. There are particular values in the moduli space where $\tau$ can be held constant (being fixed points of the monodromy), \cite{sen,DM,Sen2}:
these are $\tau=i\infty, i, e^{\frac {i\pi}3}$.
In such instances the search for BPS states can be effectively carried out, but 
the enhanced symmetries realized in this way are only a limited subset.
It is clear that for general $\tau$, things are far more complicated and the control over the BPS states is very hard to realize. Anyhow, the analysis of the constant $\tau$ examples, even though
it involved simple and far from phenomenologically interesting cases, was important to convince people that our physical intuition of the symmetry enhancement in F theory is plausible. 

\section{String junctions}

In \cite{zwie2}, using these examples, the importance of string junctions was stressed (for string junctions in F theory see \cite{zwie3,DeWolfe1,DeWolfe2,DeWolfe3,Fukae,Mohri,Choi} and also \cite{junctions}). Indeed, as is well--known, $\left(\begin{matrix}p\\ q\end{matrix}\right)$ strings may join or split and form string networks. The only condition is that the charges be conserved at the vertices. String junctions is the generic term to indicate any kind of string pattern, from elementary string prongs attached to a 7--brane  to complicated networks of strings. String junctions will be basic in the sequel.
 
Indeed, a third technique to analyze symmetry enhancement in F theory was introduced 
in \cite{zwie3}. Instead of focusing on BPS states, the idea was to consider
the lattice of string junctions related to a given system of 7--branes and define
invariant intersection numbers (scalar product) on it. Once this is done the game consists in showing that string junctions of specific composition and length 
form a realization of the root lattice of a given Lie algebra.

Before discussing how this technology works in the compactification scheme of CY fourfolds, a comment is in order to distinguish the latter from the K3 compactification. For F-theory on K3, the 7-branes are just points in the internal sphere; hence resolving singularities just amounts to separate some of those points that collapse, ending up with stacks of parallel 7-branes, possibly mutually non-perturbative. On CY fourfolds, instead, 7-branes are regarded as 4-dimensional divisors of the base space, and having stacks of parallel 7-branes after resolution is now a highly non-generic situation. In general 7-branes will intersect in many complicated ways, and, in addition, after the complete resolution of the singularity placed on codimension 1 in the base, nothing guarantees the absence of additional singularities on higher codimension loci. However, rather than attempting to control such global issues, our purpose here is more limited: we will work strictly locally, in a coordinate patch whose origin will represent the singularity, thus mimicking (locally) the situation of K3. We remark
in particular that in this way the 7--brane type (see also below) is well defined via its monodromy around the local singularity.

In the geometry of an elliptically fibered CY fourfold $X_4$, let us consider the neighborhood of a point where a group of collapsed 7--branes sits and the elliptic fiber degenerates. As we have just explained, we can limit ourselves to a neighborhood represented by the local coordinates $x,y,\sigma$. The singularity  is supposed to be located at $\sigma=0$, where $\sigma$ represents a coordinate transverse to the bunch of branes. In this sense the geometric environment is locally similar to the compactification on a K3 surface, with $\sigma$ replacing the coordinate $z$ on ${\mathbb P}^1$. The only difference is that $\sigma$ is only defined locally, while $z$ represents the full ${\mathbb P}^1$ in the familiar way. But this is sufficient for the construction we have in mind.   

For practical reasons we avoid introducing new notation and adopt that of \cite{zwie3}. As in \cite{zwie2,zwie3} we will introduce three types of 7--branes,
called $A,B$ and $C$ and summarized below:
\be
A&=&[1,0]: \quad K_A \equiv M_{1,0}^{-1}=\left(\begin{array}{lr}1&-1\\0&1\end{array}\right)\0\\
B&=&[1,-1]: \quad K_B \equiv M_{1,-1}^{-1}=\left(\begin{array}{lr}0&-1\\1&2\end{array}\right)\label{ABC}\\
C&=&[1,1]: \quad K_C \equiv M_{1,1}^{-1}=\left(\begin{array}{lr}2&-1\\1&0\end{array}\right)
\ee
$A$ represents the ordinary D7--brane. 
Every $[p,q]$-brane is characterized by the monodromy matrix $M_{p,q}$ defined by
\be
M_{p,q}= \left(\begin{matrix} 1-pq& p^2\\ -q^2 & 1+pq \end{matrix}\right)\label{Mpq}
\ee
To describe the geometry we will deform the brane configuration, by separating the branes by a slight amount, so that, afterward, all the branes will lie at different points of the $\sigma$ plane near $\sigma=0$. In order to keep track of the $SL(2,Z)$ transformation properties of the branes and strings we will draw in the $\sigma$ plane, in the neighborhood of $\sigma=0$, cuts starting from the branes and going to `infinity', where `infinity' is a conventional point where all the cuts end. For definiteness we imagine the cuts going upward. As explained in \cite{zwie2,zwie3}, an $\left(\begin{matrix}r\\ s\end{matrix}\right)$ string crossing the cut in the anticlockwise direction will
appear beyond the cut as the string $M_{p,q}^{-1} \left(\begin{matrix}r\\ s\end{matrix}\right)$.  If we drag the string down the cut through the point where the brane sits, we will have a U-dual version of the Hanany--Witten effect \cite{HW}: a third string prong
will develop, starting from the brane and joining the string in such a way that at the triple junction the charges are conserved. That is, the finite prong will have charges  $(M_{p,q}^{-1}-1) \left(\begin{matrix}r\\ s\end{matrix}\right)$.

The above are the basics about junctions. The authors of \cite{zwie3} were able to show that junctions generate a lattice. Let us consider a junction ${\bf J}$, with
endpoints on different branes and possibly at infinity. Let $b$ denote a brane index.
Then we associate to each brane the charge
\be
Q^b({\bf J}) =n_+-n_- + \sum_{k=1}^{n_b} \left| \begin{matrix} r_k&p_b \\ s_k &q_b\end{matrix}\right|\label{charge}
\ee
where $n_+$ is the number of $\left(\begin{matrix}p_b\\ q_b\end{matrix}\right)$ prongs departing from the $b$ brane, and $n_-$ is the number of $\left(\begin{matrix}p_b\\ q_b\end{matrix}\right)$ prongs ending on the $b$ brane. Moreover $n_b$, a nonnegative integer, is the number of intersections 
of ${\bf J}$ with the cut starting at the $b$ brane and $\left( \begin{matrix}r_k\\ s_k\end{matrix}\right)$ are the charges of the strings belonging to ${\bf J}$ that cross the cut at the $k$-th intersection in a counterclockwise direction. 
The charge $Q^b$ can be shown to be invariant under the cut crossing above.
Of course there is also a charge associated to the point at infinity. It will be called the {\it asymptotic} charge.

Now for a brane with label $b$ and type $[p_b,q_b]$, the outgoing   $\left(\begin{matrix}p_b\\ q_b\end{matrix}\right)$ string starting at the brane and going to infinity will be denoted ${\bf s}_b$. This is a very simple case of junction
whose charges are $Q^a({\bf s}_b)= \delta^a_b$. Moreover, given two junctions ${\bf J}_1$ and ${\bf J}_2$, their sum is naturally defined as the junction with charges
\be
Q^a({\bf J}_1+{\bf J}_2)= Q^a({\bf J}_1)+Q^a({\bf J}_2)\0
\ee

These rules define a lattice in which one can introduce a scalar product as follows: for an ${\bf s}$ elementary prong defined above we have 
\be
<{\bf s},{\bf s}>=-1,\label{prod1}
\ee
and for a three strings junction ${\bf J}_3$ we have 
\be
<{\bf J}_3,{\bf J}_3> = \left| \begin{matrix} p_i& p_{i+1}\\ q_i&q_{i+1}\end{matrix}
\right|\label{prod2}
\ee
where $i$ is an integer mod 3. It is easy to see that this definition
is independent of $i$. These rules define a (in general degenerate) metric in the junction lattice. For instance if we have $n$ branes of type $A$, one brane of type $B$ and one of type $C$ the corresponding elementary prongs $\a_i$ (i=1,\ldots,n), $\b$ and $\c$ departing from them, have the following metric:
\be
\langle \a_i,\a_j\rangle&=&-\delta_{ij}\0\\
\langle \a_i,\b\rangle&=&-1/2\0\\
\langle \a_i,\c\rangle&=&1/2\0\\
\langle \b,\b\rangle&=&-1\0\\
\langle \c,\c\rangle&=&-1\0\\
\langle \b,\c\rangle&=&1\label{metric}
\ee

Armed with these tools the authors of \cite{zwie3}, by simply selecting the junctions of given length and vanishing asymptotic charge, were able to identify the junctions
that correspond to all the roots of simply-laced Lie algebras. We will recall explicit examples below, but, especially, our purpose
will be to single out the combinations of these roots which are invariant under the symmetry (if any) of the relevant Dynkin
diagram in order to extract the roots of the corresponding  non--simply--laced Lie algebras. In this way we will construct  the root system of the  $\bf{B}_n$ and $\bf{C}_n$ series, and of $\bf{F}_4$ and $\bf{G}_2$, in terms of string junctions.  We will show in addition that all such roots can be given in terms of junctions or in terms of Jordan strings (that is, string prongs without three or higher order
string mergings). Moreover, we will interpret our results in a physical perspective in terms of branes 
and their orientifold images, fractional (involution invariant) branes and string stretching among them. 

\section{Orthogonal Lie algebras}

The $\bf{D}_n={\bf so(2n)}$ ($n\ge4$) algebras are constructed out of $n$ $A$-branes,
one $B$-brane and one $C$-brane.  The $\bf{B}_{n-1}={\bf so(2n-1)}$ ($n\ge4$) algebras are, instead, $\mathbb{Z}_2$ folding of $\bf{D}_n$ (the last two simple roots are identified) and we are going to show how this procedure will be seen by means of a resolution of type $\bf{I}^*_{n-4}$ Kodaira singularity.

\subsection{{\bf so(2n)} algebras}
Let us first review the construction of the $\bf{D}_n$ algebras, following the procedure of \cite{zwie3}.
The {\bf so(2n)} algebras are constructed with $n$ $\a$--type prongs 
$\a_i$, $i=1,\ldots,n$, a $\b$ prong and a $\c$ prong. So the relevant vector space in this case is $\mathbb{R}^{n+2}$, spanned by $\{\a_1,\ldots,\a_n,\b,\c\}$. The roots are the following:
\be
\pm(\a_i-\a_j) && 1\leq i<j\leq n\0\\
\pm(\a_i+ \a_j-\b-\c)&&1\leq i<j\leq n\label{so2nroots}
\ee
They are  $2n^2-2n$. Counting the $n$ zeroes corresponding to Cartan generators, makes $2n^2-n$, that is the dimension of the {\bf so(2n)} algebra. The meaning  of the non--zero roots is very clear in terms of strings and orientifolds. First of all, as one can see from (\ref{so2nroots}), there is no charge left at infinity by these states (no asymptotic charge). Moreover, they all have the same length (the squared norm is equal to -2, computed by means of (\ref{metric})), as it should be for simply-laced algebras. Finally, looking at the coefficients in (\ref{so2nroots}), we note that:
\begin{itemize}
\item the root $(\a_i-\a_j)$ just corresponds to the standard string stretching from the i-th $A$-brane to the j-th one;
\item the root $(\a_i+ \a_j-\b-\c)$ corresponds to a string departing from the i-th $A$-brane, going across the branch cuts of the $B$-brane and of the $C$-brane and eventually ending on the j-th $A$-brane, but with reversed orientation (so it is also departing from the j-th $A$-brane). In fact the effect of $K_CK_B$ on a fundamental string is to reverse its sign. Therefore
$C$ and $B$ can be thought of as the constituents of a non-perturbative bound state, corresponding to the \emph{orientifold} $O7^-$ of the perturbative theory of the D7's. The case $(\a_i+ \a_i-\b-\c)$ is not included because it corresponds to  non-orientable strings, which would be massive even in the collapsing limit. Therefore $(\a_i+ \a_j-\b-\c)$
junctions realize the expected antisymmetric Chan-Paton factors. On the covering space of this $\mathbb{Z}_2$-orbifold, such twisted states simply lift to strings stretching between a brane and the mirror image of another brane. This consideration allows us to write:
\begin{equation}
\bar{\a}_i\,\equiv\,\b+\c-\a_i
\label{aimage}\end{equation} 
defined as the asymptotic string departing from the orientifold image of the i-th $A$-brane. It has the correct asymptotic charge, the right squared length of a normal $\a$-prong and vanishing scalar product with $\{\a_j\}_{j\neq i}$, as it is easy to verify. In this way the root $(\a_i+ \a_j-\b-\c)$ becomes $(\a_i- \bar{\a}_j)$, thus representing the familiar string departing from the i-th brane and ending on the image of the j-th one.
\end{itemize}
All the roots we have constructed are represented by string-junctions with vanishing asymptotic charge and all have the same squared length (-2), as it should be for a simply-laced Lie Algebra.

Finally, in order to visualize the folding of the {\bf so(2n)} algebra, let us write here also its simple roots:
\be
{\bf \alpha}_i = \a_i-\a_{i+1}, \quad\quad i=1,\ldots,n-1,\quad
{\rm and} \quad {\bf\alpha}_n=\a_{n-1}-\bar\a_n\label{so2nsimpleroots}
\ee

\subsection{{\bf so(2n-1)} algebras}

As already said, these algebras are obtained from the previous ones by identifying the last two simple roots in (\ref{so2nsimpleroots}), which are exchanged by the $\mathbb{Z}_2$ outer automorphism of the {\bf so(2n)} algebra. From the point of view of the 7-branes, we can achieve this by simply identifying the last $A$-brane with the fractional one, which lies on the orientifold. So let us set $\a_0\equiv \a_n$ for the corresponding outgoing asymptotic string; the identification will thus impose the following relation:
\be
&&2\,\a_0=\b+\c\label{fract}  
\ee  
Hence the relevant vector space for the {\bf so(2n-1)} algebra is an $\mathbb{R}^{n+1}$ vector subspace of $\mathbb{R}^{n+2}$, defined by (\ref{fract}), which by the way is consistent with the fact that the fractional brane is still a D7. Notice, however, that this prong has now norm equal to 0 and also vanishing scalar product with any other vector. Thus we have to set:
\be
\langle \a_0,\a_0\rangle&=&0\0\\
\langle \a_0,\a_i\rangle&=&0\0\\
\langle \a_0,\b\rangle&=&0\0\\
\langle \a_0,\c\rangle&=&0
\ee
Some of the roots of  {\bf so(2n-1)} are represented by the junctions
\be
&\pm(\a_i-\a_j) &\quad\quad 1\leq i<j\leq n-1\0\\
& \pm(\a_i- \bar\a_j)& \quad\quad 1\leq i<j\leq n-1\label{so2n-1a}
\ee
whose physical meaning is identical to the one described in the previous section, since they just correspond to the (n-1)(2n-4) roots of the maximal {\bf so(2n-2)} subalgebra. The remaining roots are:
\be
\pm(\a_i-\a_0)&\approx&\pm(\a_i-\bar\a_0) \quad\quad i=1,\ldots n-1\label{so2n-1b}
\ee
These correspond instead to strings stretching from the $A$-branes to the fractional brane sitting on top of the orientifold (both the orientations are possible). The equivalence is due to the invariance of the fractional brane under the orientifold involution, which means, as stated in (\ref{fract}), $\a_0=\bar\a_0$ for the corresponding asymptotic string. A further comment is in order: due to the vanishing norm of the fractional brane, these states have now squared length equal to -1! It is clear then that they correspond to the short roots of the non-simply laced algebra $\textbf{B}_{n-1}$. Altogether these are $(n-1)(n-2)+(n-1)(n-2)+2(n-1)=(n-1)(2n-2)$ non-zero roots.  They fill up the root set of {\bf so(2n-1)}. Counting $n-1$ zeroes corresponding to the Cartan subalgebra this yields the dimension of  {\bf so(2n-1)}.

The simple roots of {\bf so(2n-1)} are:
\be
\alpha_i=\a_i-\a_{i+1}, \quad\quad i=1,\ldots, n-2,\quad {\rm and}
\quad \alpha_{n-1}= \a_{n-1} -\a_0 \label{simplerootsso2n-1}
\ee
Therefore the roots $\alpha_i$,  $i=1,\ldots, n-2$ are long, while $\alpha_{n-1}$ is short.
All in all, in this Lie Algebra there are 2n-2 short roots, while the remaining ones are long, and all are still represented by string junctions with vanishing charge at infinity.

Actually we can say more. The physical meaning of the roots (\ref{so2n-1a}) and (\ref{so2n-1b}) will tell us their behavior under the breaking of the odd orthogonal gauge algebra to the maximal subalgebra that can be realized perturbatively. 
Suppose we resolve the non-split $\bf{I}^*_{n-4}$ Kodaira singularity, the one relevant for the $\bf{B}_{n-1}$ algebra, in two groups of 7-branes, one made of n-1 $A$-branes on top of each other, and the other made by the fractional $A$-brane on top of the $CB$ orientifold. In this way, the manifest \emph{perturbative} subalgebra of {\bf so(2n-1)} will be ${\bf su(n-1)}\times {\bf u(1)}$. Hence, for the breaking
\be
{\bf so(2n-1)}&\longrightarrow&{\bf su(n-1)}\times \bf{u(1)}
\ee 
the branching rule for the adjoint representation is\footnote{We will not keep track of the {\bf u(1)} charges, as they cannot be detected by an analysis like ours based on string junctions.}:

\be
(n-1)(2n-1)&\longrightarrow&(n-1)^2-1+1+2\times\frac{(n-1)(n-2)}{2}+2\times(n-1)\0\\
\ee
that is, the adjoint of {\bf so(2n-1)} goes into the adjoint plus two copies of the 2-antisymmetric plus two copies of the fundamental of  ${\bf su(n-1)}\times \bf{u(1)}$.\\
It is very easy now to match this representation content with the roots (\ref{so2n-1a}), (\ref{so2n-1b}). 
\begin{itemize}
\item The first set of roots in (\ref{so2n-1a}) (and the $n-1$ zeroes corresponding to the Cartan generators) fill up the weights of the $(n-1)^2$-dimensional adjoint representation of  ${\bf su(n-1)}\times \bf{u(1)}$, i.e. they correspond to the gauge vectors of the manifest perturbative subalgebra. 
\item The second set of roots in (\ref{so2n-1a}) fill two copies of the 2-antisymmetric representation\footnote{More precisely, the 2-antisymmetric and the (n-2)-antisymmetric, since their string representatives have opposite orientations.} of ${\bf su(n-1)}$, and are therefore responsible of the enhancing of the perturbative subalgebra ${\bf su(n-1)}\times \bf{u(1)}$ to the maximal subalgebra {\bf so(2n-2)}.
\item The roots in (\ref{so2n-1b}) fill up  two copies of the fundamental\footnote{More precisely, the fundamental and the (n-1)-antisymmetric (antifundamental), since their string representatives have opposite orientations.} of ${\bf su(n-1)}$, since, as said, they are just the strings stretched between the fractional brane and one of the n-1 $A$-branes in the stack.   
\end{itemize}

\section{$E_6$ and $F_4$}

We want now to make an analogous construction that leads from a 7--brane model for $E_6$  to the one corresponding to $F_4$, since the latter algebra can be viewed as the folding of the former one under the $\mathbb{Z}_2$ automorphism group of its Dynkin diagram.
Let us start by reviewing the procedure for $E_6$,  following again \cite{zwie3}.

\subsection{The $E_6$ algebra}
$E_6$ is constructed out of 5 $A$-branes, one $B$-brane and 2 $C$-branes. Hence the string realization of the $E_6$ algebra is based on 5 prongs $\a_1,\ldots,\a_5$, one prong $\b$ and two prongs $\c_1,\c_2$.  The non--zero roots are identified with the junctions
\be
&\pm(\a_i-\a_j),& \quad\quad 1\leq i<j\leq 5 \0\\
& \pm(\a_i-\a_j-\b-\c_k),& \quad\quad 1\leq i<j\leq 5, \quad\quad k=1,2\label{E6a} 
\ee
which are 20+40=60, together with the junctions 
\be
\pm(\sum_{k=1}^5 \a_k -\a_i-2\b-\c_1-\c_2), \quad\quad 1\leq i\leq 5\label{E6b}
\ee
and
\be
\pm(\c_2-\c_2) \label{E6c}
\ee
Altogether they make 72 roots (to be added to the 6 zero eigenvalues due to the Cartan generators). These are all the junctions
with square length -2 and vanishing asymptotic charges. 

In order to visualize the folding of the $E_6$ algebra, let us write here its simple roots:
\be
\alpha_1=\a_1-\a_2, &\quad \alpha_2=\a_2-\a_3, &\quad\alpha_3=\a_3-\a_4,\0\\
\alpha_4=\a_4-\bar{\a}^1_5,&\quad\alpha_5=\c_1-\c_2,&\quad
\alpha_6=\a_4-\a_5\label{E6simpleroots}
\ee
An {\bf so(10)} subalgebra with simple roots $\{\alpha_i\}_{i\neq5}$ is manifest.

Let us make a comment concerning the string interpretation of this construction.
We start by defining the images of the ${\bf a}$-prongs as follows:
\be
&\bar{\a}_i^I\,\equiv\,\b+\c_I-\a_i&\quad\quad 1\leq i\leq5\quad \textrm{and}\quad I=1,2
\label{aImage}\ee
according to which each of the two $C$-branes is taken to form an orientifold with the $B$-brane. Using the same scalar product (\ref{metric}) in  $\mathbb{R}^8$ generated by  $\{\a_1,\ldots,\a_5,\b,\c_1,\c_2\}$, with the addition of $\langle \c_I,\c_J\rangle=-\delta_{ij}$, one can see that the definition (\ref{aImage}) is still compatible with the metric behavior of the $\a$-prongs of (\ref{aimage}).  The second set of junctions in (\ref{E6a}) can be rewritten in the by now familiar way
\be
\pm(\a_i-\bar {\a}_j^I)&& \quad\quad 1\leq i<j\leq 5,\quad I=1,2\label{E6a'}
\ee

\subsection{The $F_4$ algebra}

$F_4$ is algebraically generated by folding the $E_6$ Dynkin diagram under its $\mathbb{Z}_2$ symmetry group. Acting on the simple roots in (\ref{E6simpleroots}), this symmetry maps $\alpha_1\to \alpha_5$, $\alpha_2\to \alpha_4$, while leaving $\alpha_3$ and $\alpha_6$ unchanged. In terms of F--strings, this is generated by the prong correspondences
\be
\a_1 &\longrightarrow& \a_3+\a_4 +\a_5-\b-\c_2\0 \label{F41}\\
\a_2 &\longrightarrow& \a_3+\a_4 +\a_5-\b-\c_1\0\\
\c_1 &\longrightarrow&\a_3+\a_4 +\a_5-\b-\a_2 \0\\
\c_2 &\longrightarrow&\a_3+\a_4 +\a_5-\b-\a_1
\ee
while $\a_3,\a_4,\a_5,\b$ remain unchanged. 

The junctions invariant under these transformations are
\be
&\pm(\a_3-\a_4),&\quad\quad \pm(\a_1+\a_3-\b-\c_2),\0\\
&\pm(\a_4-\a_5), &\quad\quad\pm( \a_1+\a_4-\b-\c_2),\0\\
 &\pm(\a_3-\a_5),&\quad\quad\pm( \a_1+\a_5-\b-\c_2),\0\\
&\pm(\a_1+\a_2+\a_4+\a_5-2\,\b-\c_1-\c_2),
&\quad\quad \pm(\a_2+\a_3-\b-\c_1),\0\\ 
 &\pm(\a_1+\a_2+\a_3+\a_5-2\,\b-\c_1-\c_2),
&\quad\quad\pm(\a_2+\a_4-\b-\c_1),\0\\ 
&\pm(\a_1+\a_2+\a_3+\a_4-2\,\b-\c_1-\c_2),
&\quad\quad \pm(\a_2+\a_5-\b-\c_1) \label{F42}
\ee
In addition the linear combinations of ${\bf E}_6$ roots which are invariant are
\be
\pm(\a_1-\a_3+\a_4+\a_5-\b-\c_2), \qquad \pm(\a_2-\a_3+\a_4+\a_5-\b-\c_1),&&\0\\
\pm(\a_1+\a_3-\a_4+\a_5-\b-\c_2), \qquad\pm(\a_2+\a_3-\a_4+\a_5-\b-\c_1),&&\0\\
\pm(\a_1+\a_3+\a_4-\a_5-\b-\c_2), \qquad\pm(\a_2+\a_3+\a_4-\a_5-\b-\c_1),&& \0\\
\pm(\a_1-\a_2+\c_1-\c_2),\,\;\,\quad\quad\qquad\pm(\a_1+\a_2+2\,\a_3-2\,\b-\c_1-\c_2),\0\\ 
\pm(\a_1+\a_2+2\,\a_4-2\,\b-\c_1-\c_2), &&\0\\ 
\pm(\a_1+\a_2+2\,\a_5-2\,\b-\c_1-\c_2),&& \0\\ 
\pm( 2\,\a_1+\a_2+\a_3+\a_4+\a_5-3\,\b-\c_1-2\,\c_2),&&\0\\
\pm(\a_1+2\,\a_2+\a_3+\a_4+\a_5-3\,\b -2\,\c_1-\c_2).&&\label{F43}
\ee 
All in all we have 24 short + 24 long = 48 roots, still represented by string junctions with vanishing charge at infinity. Adding the four zeros corresponding to the Cartan generators yields a total of 52, the dimension of ${\bf F_4}$. 

It is not hard to single out a set of simple roots for the set (\ref{F42},\ref{F43}):
\be
&\alpha_1=\a_1-\a_2+\c_1-\c_2 , &\quad \alpha_2= \a_2-\a_3+\a_4-\bar\a_5^1 ,\0\\
&\quad\quad \alpha_3= \a_3-\a_4 ,&\quad\quad \alpha_4= \a_4-\a_5 .\label{F4simpleroots}
\ee
The first two are long (squared length -4), the last two short (squared length -2). Using the scalar product (\ref{metric}) with one more $\c$-prong (with $\langle \c_I,\c_J\rangle=-\delta_{ij}$) and the definition (\ref{aImage}), we get for the nonvanishing Cartan matrix
elements 
\be
&\langle \alpha_1,\alpha_1\rangle = \langle \alpha_2,\alpha_2\rangle=-4, 
&\quad\quad \langle \alpha_3,\alpha_3\rangle = \langle \alpha_4,\alpha_4\rangle =-2,\0\\
&\langle \alpha_1,\alpha_2\rangle = \langle \alpha_2,\alpha_3\rangle=2, 
&\quad\quad \langle \alpha_3,\alpha_4\rangle =1\label{F4length}
\ee
By comparing with the simple roots of ${\bf E}_6$, we see that
$\alpha_3$ and $\alpha_6$ become the short simple roots of $F_4$, since they are left unchanged by the $\mathbb{Z}_2$ symmetry;  as long simple roots of $F_4$, instead, we take the two invariant combinations out of the remaining four simple roots of $E_6$ that are pairwise exchanged by $\mathbb{Z}_2$: these are clearly $\alpha_1+\alpha_5$ and $\alpha_2+\alpha_4$. 

With the labeling (\ref{F4simpleroots}) the roots (\ref{F42}) and (\ref{F43}) coincide with the roots of \cite{Cornwell}, vol. II, App.F, ch.8.
For instance, for the very last one in (\ref{F43}), one gets
\be
\a_1+2\,\a_2+\a_3+\a_4+\a_5-3\,\b -2\,\c_1-\c_2&=& 
2\alpha_1+4\alpha_2+3\alpha_3+\alpha_4\0 \label{lastroot}
\ee

To summarize, in order to write down the string-junctions representing the roots of the non-simply laced algebra just constructed, we have proceeded in two steps:

\begin{itemize}
\item we have singled out the roots of the parent ${\bf E_6}$ algebra that are not touched by the $\mathbb{Z}_2$ symmetry, which, therefore, remain with the same square length: these are the analogs of the roots (\ref{so2n-1a}) of the manifest {\bf so(2n-2)} subalgebra of {\bf so(2n-1)}; 
\item to find the remaining roots, which therefore have double the length of the previous ones, we have built up singlets under the $\mathbb{Z}_2$ symmetry, by taking linear combinations of the vectors.
\end{itemize}

\begin{figure}[htbp]
    \hspace{-0.5cm}
\begin{center}
    \includegraphics[scale=0.5]{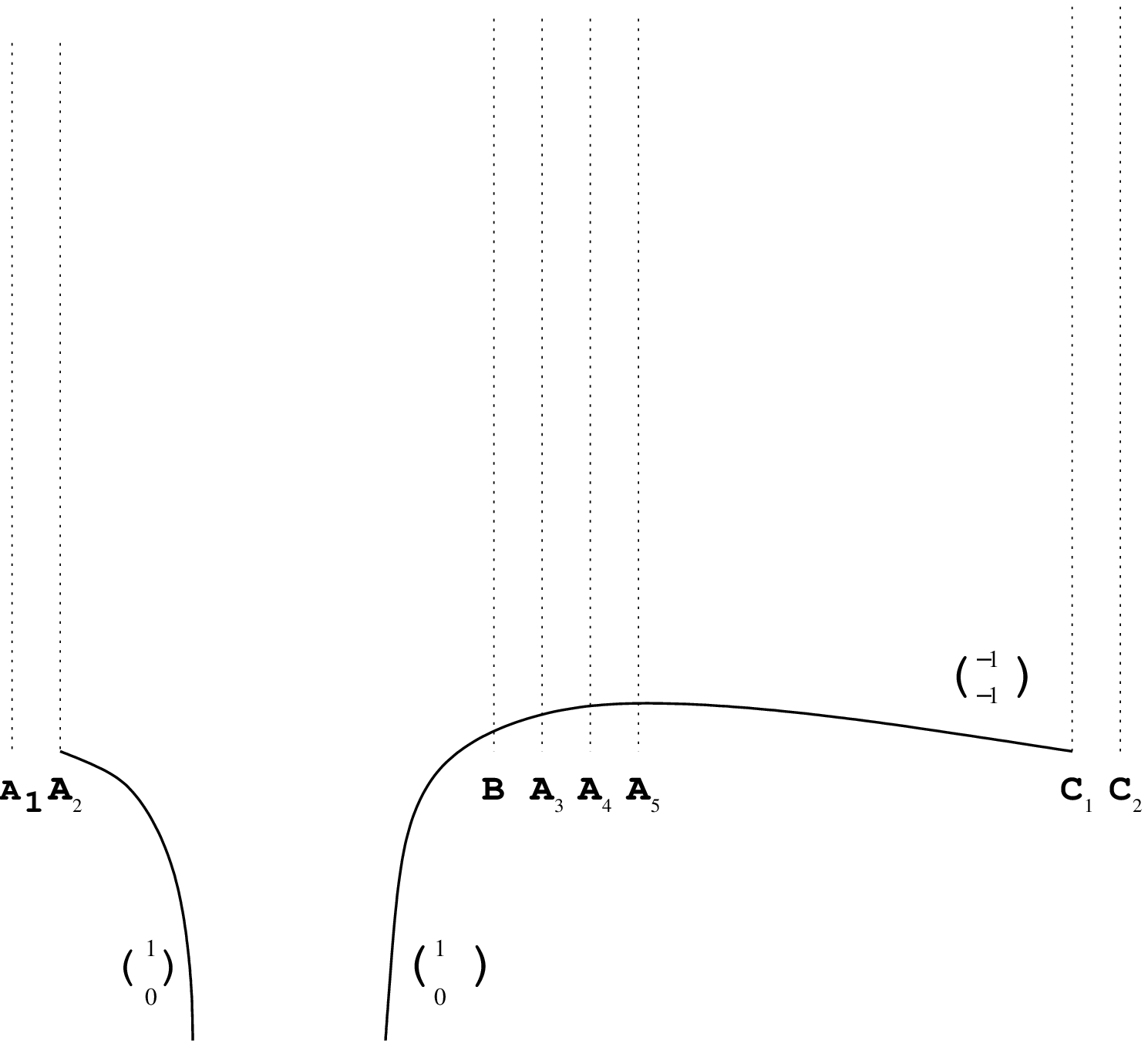}
    \end{center}
\caption{\emph{\small  The Jordan strings representing the  junctions $\a_2$ and $\a_3+\a_4+\a_5-\b-\c_1$}}
    \label{fig:F1}
\end{figure}

What we have said so far simply means that the roots of the Lie algebra ${\bf F_4}$ can be constructed in terms of 
junctions, i.e. the folding of ${\bf E_6}$ leads again to string junctions. It remains for us to understand the origin
of the ${\mathbb Z}_2$ symmetry of  ${\bf E_6}$. To this end we have to unravel the meaning of the transformations (\ref{F41}).
We will resolve the $E_6$ singularity ($IV^{*s}$, in Kodaira classification) by arranging our 8 branes, for instance, as follows: we take a group formed by $BA_3A_4A_5$ at the center, then $A_1A_2$ on the left and $C_1C_2$ on the right, with the relevant cuts
going upward. Looking at the first of (\ref{F41}) the $\a_1$ on the left is just the usual elementary prong going downward to infinity. 
The junction on the right ($\a_3+\a_4 +\a_5-\b-\c_2$) is also going to infinity and its asymptotic is the same as $\a_1$. This junction can be easily undone and represented by a Jordan string that ends on  $C_2$ coming from the left, after having crossed the cuts of $B, A_5,A_4$ and $A_3$. This is a Jordan string that, after the crossings, has the charge of a fundamental string.
Indeed one can easily verify that $K_B K_A^3 \left(\begin{matrix} -1\\ -1 \end{matrix}\right) =  \left(\begin{matrix} 1\\0 \end{matrix}\right)$.
In other words, looking from the left through the screen formed by $BA_3A_4A_5$ at a string ending on $C_2$, one sees a fundamental string.  A similar construction holds for the second transformation in (\ref{F41}) with $A_1,C_2$ exchanged with $A_2,C_1$ (see Fig.\ref{fig:F1}).

\begin{figure}[htbp]
    \hspace{-0.5cm}
\begin{center}
    \includegraphics[scale=0.5]{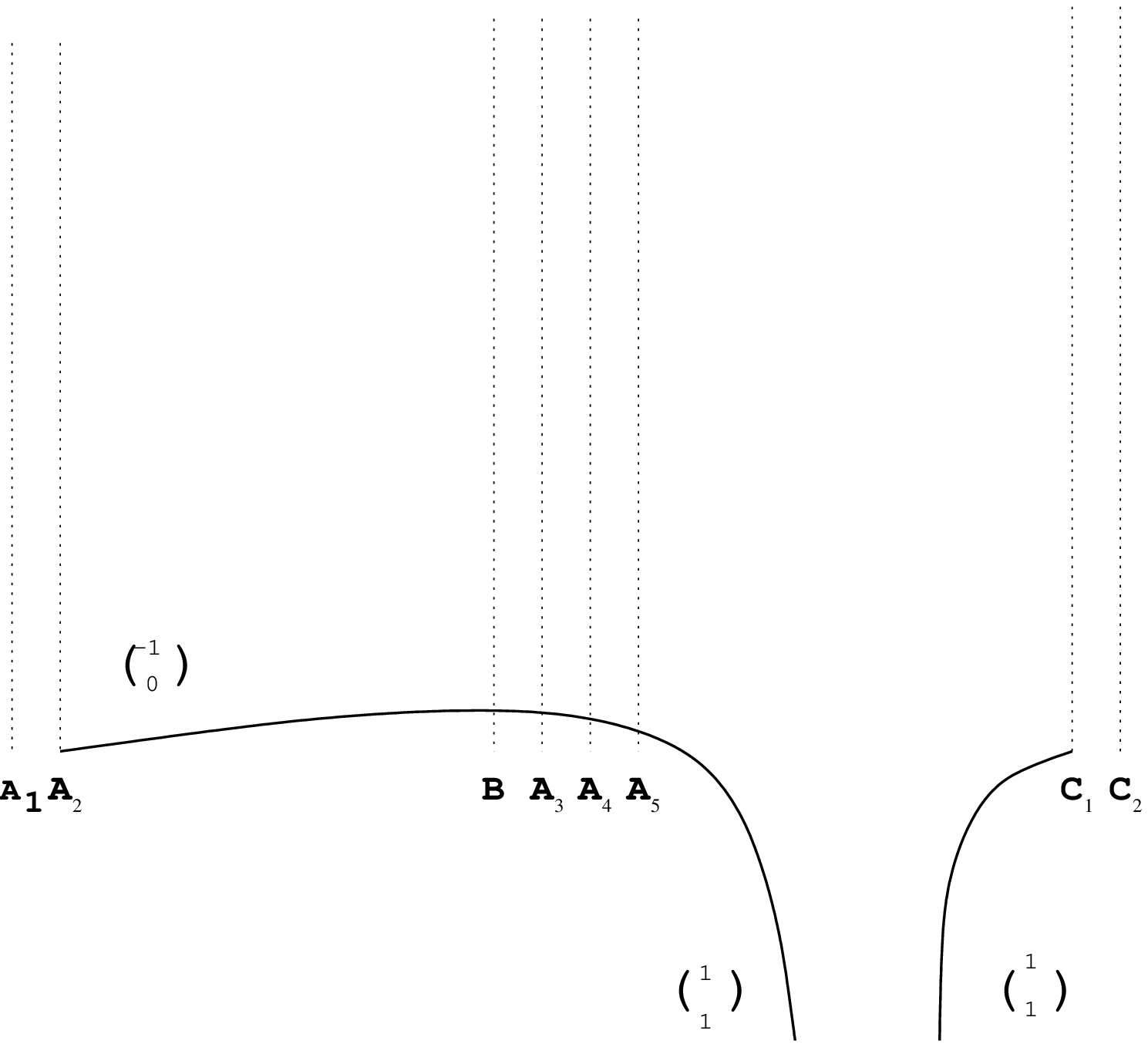}
    \end{center}
\caption{\emph{\small  The Jordan strings representing the  junctions $\c_1$ and $\a_3+\a_4+\a_5-\b-\a_2$}}
    \label{fig:F2}
\end{figure}

Let us consider next the third transformation in (\ref{F41}). In this case the $\c_1$ prong on the left is the usual elementary prong
departing from $C_1$ and going down to infinity. The junction $\a_3+\a_4 +\a_5-\b-\a_2$ on the right can be undone and replaced
by a string ending on $A_2$ and crossing backward successively the cuts of $B,A_3,A_4,A_5$, and emerging behind the  
$BA_3A_4A_5$ screen as a $\c$ prong that goes down to infinity. In other words looking from the right through the  $BA_3A_4A_5$ screen one sees $c$ strings instead of the original (oppositely oriented) fundamental strings (see Fig.\ref{fig:F2}).  Likewise for the fourth transformation in (\ref{F41}) with $A_2,C_1$ exchanged with $A_1,C_2$. The conclusion is that the screen formed by $BA_3A_4A_5$ changes fundamental strings to $\c$ strings while reversing the orientation, and viceversa. The fact that fundamental strings can be seen as 
oppositely oriented $\c$ strings and viceversa, creates a ${\mathbb Z}_2$ symmetry among the roots of ${\bf E}_6$. This symmetry is only evident when $B, A_3,A_,A_5$ collapse before the others branes, and $A_1,A_2$ and $C_1,C_2$ collapse symmetrically with respect to  the  $BA_3A_4A_5$ screen. The orbifold with respect to this ${\mathbb Z}_2$ symmetry gives rise, in the collapsing limit,
to ${\bf F_4}$. This is our F-string description of the ${\bf E}_6$ folding to  ${\bf F_4}$.

\section{The ${\bf G_2}$ algebra}

Let us now carry out the same procedure for the ${\bf G_2}$ algebra that comes from the {\bf so(8)} one via a triple folding under the extended outer automorphism group of  $\bf{D}_4$ (due to its triality). Thus we briefly review the root structure of this parent algebra.
As we saw in the first section, the {\bf so(8)} algebra is constructed with $4$ $\a$-prongs, one $\b$-prong and one $\c$-prong. The relevant vector space is an $\mathbb{R}^6$ generated by $\{\a_1,\ldots,\a_4,\b,\c\}$ and the roots are:
\be
&\pm(\a_i-\a_j),&\quad\quad 1\leq i<j\leq 4\0\\
& \pm(\a_i+\a_j-\b-\c),& \quad\quad 1\leq i<j\leq 4\label{so8roots}
\ee
They are  24. Adding the 4 Cartan generators makes 28 dimensions.\\
The simple roots are:
\be
&\alpha_1 = \a_1-\a_{2}, &\quad\quad  \alpha_2= \a_2-\a_3,\0\\  
&\alpha_3= \a_3-\a_4,& \quad \quad \alpha_4=\a_3+\a_4 -\b-\c.\label{so8simpleroots}
\ee
The symmetries of the $\bf{D}_4$ Dynkin diagram are the ones of the equilateral triangle, namely they form a group $\bf{T}^3$ made of the $\mathbb{Z}_3$ rotations of the roots $\alpha_{1,3,4}$ , together with the three reflections:
\be
\tau_1(\alpha_{1,4})=\alpha_{4,1},&\quad \tau_1(\alpha_i)=\alpha_i,&\quad i=2,3\0\\
\tau_2(\alpha_{3,4})=\alpha_{4,3},&\quad\tau_2(\alpha_i)=\alpha_i,&\quad i=1,2\label{tau}\\
\tau_3(\alpha_{1,3})=\alpha_{3,1},&\quad\tau_3(\alpha_i)=\alpha_i,&\quad i=2,4\0
\ee
For instance, by folding $\bf{D}_4$ under $\tau_2$ alone we obtained in the first section the algebra $\bf{B}_3$, corresponding to $\bf{so(7)}$. As far as ${\bf G_2}$ is concerned, instead, we need all the reflections (actually just two of them will be enough as we are going to show), but we can disregard the invariance under the rotations, since the latter are simply products of two reflections.

Hence, in terms of string junctions the reflections (\ref{tau}) are generated by
\be
\tau_1(\a_1)&=& \frac 12 (\a_1+\a_2+\a_3+\a_4-\b-\c)\0\\
\tau_1(\a_2)&=& \frac 12 (\a_1+\a_2-\a_3-\a_4+\b+\c)\0\\
\tau_1(\a_3)&=& \frac 12 (\a_1-\a_2+\a_3-\a_4+\b+\c)\0\\
\tau_1(\a_4)&=& \frac 12 (\a_1-\a_2-\a_3+\a_4+\b+\c)\label{tau1}
\ee
and

\be
\tau_3(\a_1)&=& \frac 12 (\a_1+\a_2+\a_3-\a_4)\0\\
\tau_3(\a_2)&=&\frac 12 (\a_1+\a_2-\a_3+\a_4)\0\\
\tau_3(\a_3)&=& \frac 12 (\a_1-\a_2+\a_3+\a_4)\0\\
\tau_3(\a_4)&=& \frac 12 (-\a_1+\a_2+\a_3+\a_4)\label{tau3}
\ee
while, as we already know (compare with (\ref{fract})),

\be
\tau_2(\a_4)=\b+\c-\a_4,\quad\quad \tau_2(\a_i)=\a_i,\quad i=1,2,3\label{tau2}
\ee
and in any case $\b$ and $\c$ are left unchanged

\be
\tau_i(\b)=\b,\quad\quad \tau_i(\c)=\c, \quad\quad i=1,2,3\0
\ee

First of all, notice that only two independent constraints on $\mathbb{R}^6$ are imposed by the joint action of these three reflections, which is consistent with the rank being lowered by two units. Indeed, using the usual definition for the images (\ref{aimage}), the correspondences (\ref{tau1}), (\ref{tau3}) and (\ref{tau2}) amount to the following identifications:

\be
\tau_1&\Longrightarrow&\bar\a_4\,\approx\,\a_2+\a_3-\a_1\0\\
\tau_2&\Longrightarrow&\a_4\,\approx\,\bar\a_4\0\\
\tau_3&\Longrightarrow&\a_4\,\approx\,\a_2+\a_3-\a_1
\ee
We soon recognize in the second constraint above the fractional nature of the forth $A$-brane and we immediately see that one of the three identifications is not independent of the other two. Thus, the relevant vector space for the $\bf{G}_2$ algebra will be given by the following quotient:
 
\be
\frac{\textrm{Span}\,\{\a_1,\ldots,\a_4,\b,\c\}}{\{\a_4\approx\bar\a_4\approx\a_2+\a_3-\a_1\}}&\simeq&\mathbb{R}^4
\ee

As in the previous cases, let us now proceed to the explicit construction of the roots.\\
By looking at the 4 simple roots of {\bf so(8)} we readily notice that just one of them, $\alpha_2$ is not touched at all by any of the elements of the triality group $\bf{T}^3$ (as it corresponds to the middle node in the $\bf{D}_4$ Dynkin diagram): thus it corresponds to a short root and it passes to the quotient keeping its squared length equal to -2. The remaining three simple roots of {\bf so(8)} are pairwise exchanged by the $\{\tau_i\}_{i=1,2,3}$, so that there exist clearly only one invariant linear combination of them, namely $\alpha_1+\alpha_3+\alpha_4$: this corresponds to a long root and, as such, it survives to the quotient but it has three times the squared length of the previous one, i.e. -6. Hence, the simple roots of $\bf{G}_2$ will be:

\be
\beta_1\,\equiv \,\a_1-\a_2+2\a_3-\b-\c &&\quad\quad \beta_2\,\equiv\, \a_2-\a_3\label{G2simpleroots}
\ee
Using the usual scalar product (\ref{metric}), it is easy to find out the Cartan matrix of the $\bf{G}_2$ algebra:

\be
\langle \beta_1,\beta_1\rangle =-6, \quad\quad \langle \beta_2,\beta_2\rangle =-2,
\quad\quad \langle \beta_1,\beta_2\rangle =3\0
\ee

We are now ready to write down explicitly all the roots of $\bf{G}_2$. As we have seen for the simple roots, of the 24 roots of the parent $\bf{D}_4$ one fourth of them passes directly to the quotient without any change (short roots):

\be
\pm(\a_2-\a_3),&\qquad\pm(\a_1-\bar\a_2),&\qquad\pm(\a_1-\bar\a_3)
\ee
They are 6 and explicitly look like:

\be
\pm\beta_2&=&\pm(\a_2-\a_3),\0\\
\pm(\beta_1+2\beta_2)&=&\pm( \a_1+\a_2-\b-\c),\\
\pm(\beta_1+\beta_2)&=& \pm(\a_1+\a_3-\b-\c).\0
\ee
Of the remaining three fourths of the $\bf{D}_4$ roots, only one third survives the quotient, namely the 6 singlet combinations (long roots), and we will write them directly in the explicit form:

\be
\pm\beta_1&=&\pm(\a_1-\a_2+2\a_3-\b-\c),\0\\
\pm(\beta_1+3\beta_2)&=&\pm(\a_1+2\a_2-\a_3-\b-\c),\label{G2posroots}\\
\pm(2\beta_1+3\beta_2)&=&\pm(2\a_1+\a_2+\a_3-2\b-2\c).\0
\ee

All in all we have 6 short + 6 long = 12 roots, still represented by string junctions with vanishing charge at infinity. Adding the 2 zeroes corresponding to the Cartan generators makes a total of 14, the dimension of $\bf{G}_2$.

Notice that the $\a_4$-prong has disappeared from the roots of $\bf G_2$. However this is only apparent. In fact, in the simple root $\beta_1$ the junction $\a_1-\a_2$ is easily interpretable, but the junction $2\a_3-\b-\c$ would represent an $A$-string starting from the brane $A_3$, circling around the $CB$ orientifold and returning to the \emph{same} brane with \emph{opposite} orientation, i.e. it would be a non-orientable string. Such a string would be massive in the collapsing limit. The paradox is explained by correctly interpreting  $\beta_1$ as $\a_1-\a_2 + \a_3-\a_4+ \a_3-\bar\a_4$. In this case, all
the involved junctions ($\a_1-\a_2$, $\a_3-\a_4$ and $\a_3-\bar\a_4$) are massless Jordan strings,  and the paradox disappears.
 
\begin{figure}[htbp]
    \hspace{-0.5cm}
\begin{center}
    \includegraphics[scale=0.5]{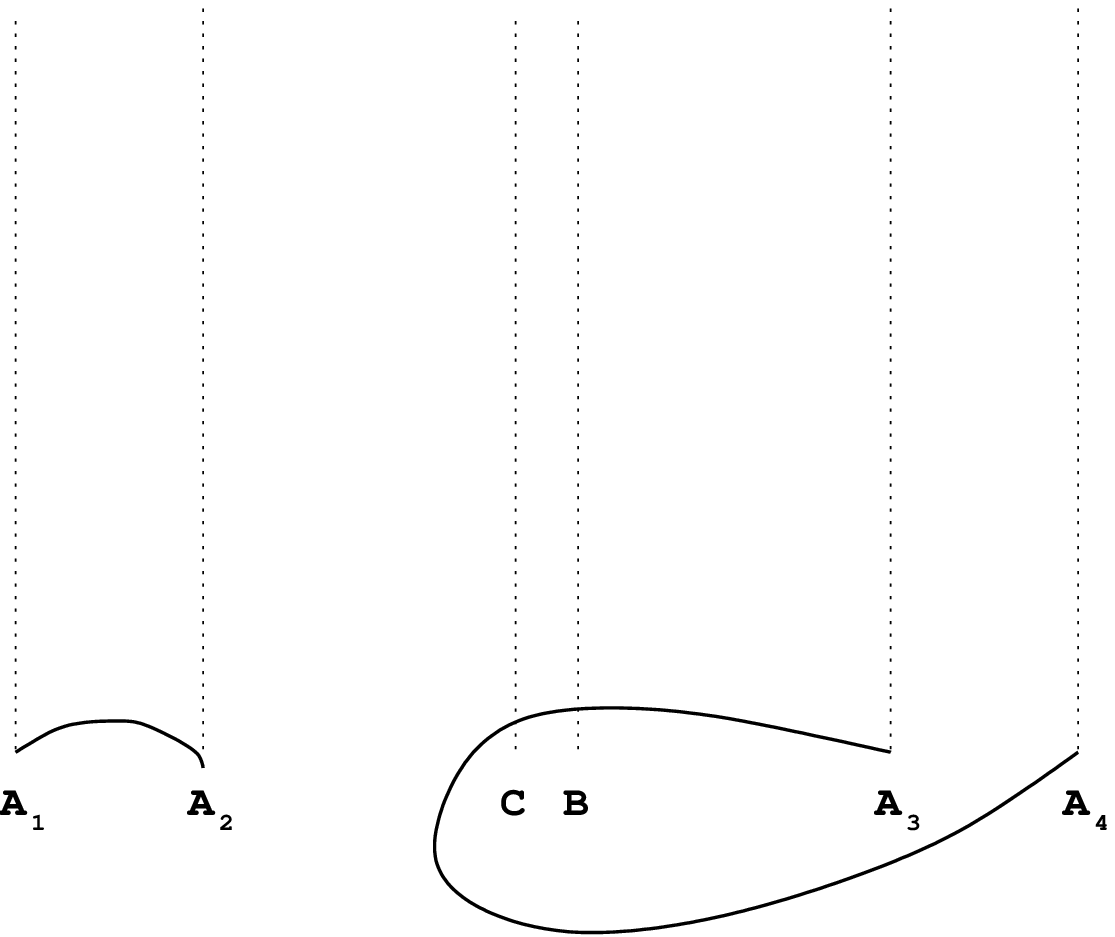}
    \end{center}
\caption{\emph{\small  The Jordan strings representing the junctions $\a_1-\a_2$ and $\a_3+\a_4-\b-\c$.}}
    \label{fig:F3}
\end{figure}

So far we have shown that the folding of the ${\bf D}_4$ algebra down to ${\bf G}_2$ can be implemented in terms of string junctions.
We would like now to show that the symmetry responsible for such folding can be interpreted in a natural way as a symmetry
of F-string configurations. The transformation (\ref{tau2}) has already been understood by interpreting a $CB$ bound state as
an orientifold. This reduces the transformations that need to be interpreted to the set (\ref{tau1}). It is easy to see that
these transformations can be replaced by the following ones
\be
&&\tau_1(\a_1+\a_2)=\a_1+\a_2\0\\
&&\tau_1(\a_1-\a_2)=  \a_3+\a_4-\b-\c \0\\
&&\tau_1(\a_1-\a_3)=  \a_2+\a_4-\b-\c \0\\
&&\tau_1(\a_3-\a_4)=  \a_3-\a_4 \label{tau1'}
\ee
In fact from these we can derive (\ref{tau1}) and viceversa. The interpretation of the first and last equations are of course trivial.
As for the others let us consider the following brane  resolution of the relevant $I_0^{*s}$ singularity. The $CB$ block at the center, $A_1,A_2$ at the left and $A_3,A_4$ at the right. Then $\a_1-\a_2$ represents a fundamental string departing from $A_1$ and ending on $A_2$, while $ \a_3+\a_4-\b-\c$  represents a fundamental string departing from $A_3$ going around the orbifold $CB$ and returning to $A_4$ with reversed orientation. We know that the latter is the junction $\a_3-\bar \a_4$ which has been already identified by $\tau_2$ with 
 $\a_3-\a_4$, see  Fig.\ref{fig:F3}. The symmetry of this configuration under reflection with respect to the $CB$ block is evident (in this local representation).

Going to the third of the (\ref{tau1'}), the junction $\a_1-\a_3$ is a fundamental string that departs from $A_1$ and ends on $A_3$ without crossing any cut. On the other hand $ \a_2+\a_4-\b-\c$ is a fundamental string that departs from $A_2$, crosses the $C$ and $B$  cuts 
and ends on $A_4$ with reverse orientation.  Again, due to the presence of an orientifold we identify $A_4$ with its mirror image, which means identifying  $ \a_2+\a_4-\b-\c$ with a string stretching from $A_2$ to $A_4$ without crossing cuts. Even after these moves the symmetry of the configuration is not immediately evident. But it is easy to see that using (\ref{tau1'}) and in addition
$\tau_1(\a_2-\a_3)= \a_2-\a_3$, which also follows from (\ref{tau1}), one can pass from the second to the third transformation in (\ref{tau1'}). Since the former is a symmetry also the latter is.

\section{Symplectic algebras}

The last non-simply laced algebras in the classification are the symplectic ones. 
The {\bf sp(n)} ($n>1$) algebra\footnote{We adopt here the convention for which $n$ stands for the rank of the algebra. So $sp(2)\sim so(5)$.} is a $\mathbb{Z}_2$ folding of the ${\bf su(2n)}$ algebra under its $\mathbb{Z}_2$ outer automorphism that reflects the nodes of the  ${\bf A_{2n-1}}$ Dynkin diagram with respect to the central one.
The ${\bf A_{2n-1}}$ algebra is obviously realized by means of $\a$-type junctions stretching among $2n$  D7 branes on top of one another (giving rise to $U(2n)$ gauge group). The positive roots of this algebra are: 
\be
\a_i-\a_j,&& \quad\qquad 1\leq i<j\leq 2n \label{Aposroots}
\ee
while the simple ones are:
\be
\alpha_i\,\equiv\,\a_i-\a_{i+1}, &&\quad\quad 1\leq i\leq 2n-1 \label{Asimproots}
\ee
The $\mathbb{Z}_2$ symmetry acts on these simple roots as:
\be
\alpha_i& \leftrightarrow& \alpha_{2n-i}\label{Z2alpha}
\ee
so that $\alpha_n$ (corresponding to the central node) remains unchanged. This $\mathbb{Z}_2$ symmetry is realized by the correspondences
\be
\a_i &\leftrightarrow& -\a_{2n-i+1}\label{Z2}
\ee
As is evident from (\ref{Z2}), by imposing such constraints we half the dimension $\mathbb{R}^{2n}$ vector space we started with, so that the relevant vector space for the {\bf sp(n)} algebra will be the following quotient:
\be
\frac{\textrm{Span}\,\{\a_1,\ldots,\a_{2n}\}}{\{\a_i+\a_{2n-i+1}\approx 0\}_{i=1,\ldots,n}}&\simeq&\mathbb{R}^n
\ee

Following the example of the previous cases it is easy to explicitly construct the positive roots. First we have
\be
2\a_i-2\a_{2n-i+1},&&\quad\quad 1\leq i\leq n\label{spnroots1}
\ee
that descend straight from the invariant roots of ${\bf su(2n)}$. They are long roots. The other positive roots are, as usual, invariant
combination of the  ${\bf su(2n)}$ ones. They are
\be
\a_i-\a_j+\a_{2n-j+1}-\a_{2n-i+1}, &&\quad \quad 1\leq i<j<2n,\quad i+j\leq 2n\label{spnroots2}
\ee
They are $n(n-1)$ short positive roots.  It is easy to see that a set of simple roots is
\be
\gamma_i&\equiv&\a_i-\a_{i+1}+\a_{2n-i}-\a_{2n-i+1}, \quad\qquad 1\leq i\leq n-1\0\\
\gamma_n&\equiv&2\a_n-2\a_{n+1}.
\ee
These are $n-1$ short and 1 long. The Cartan matrix of ${\bf C_n}$ is easily recovered, using the scalar product (\ref{metric}) .

In conclusion it is easy to realize the folding of  ${\bf A_{2n-1}}$ and obtain the roots of the Lie algebra ${\bf C}_n$ in terms 
of junctions. But this is only a formal operation, without any string interpretation behind it. In fact, having at hand only 
$A$ branes it is impossible to construct an orientifold or a screen like in the previous cases, since 
fundamental strings that cross $A$--cuts remain fundamental strings. Thus, being not aware of any realization of $O7^+$ planes out of F theory 7-branes\footnote{Without intersecting branes, Sp(n) gauge symmetry can be realized only in the presence of $O7^+$ planes.}, we conclude that the ${\bf C_n}$ Lie algebras cannot be realized in the geometry considered in this paper. 

%%%%%%%%%%%%%%%%%%%%
\acknowledgments
%%%%%%%%%%%%%%%%%%%

We would like to thank Sergio Cecotti for his course on F theory.
L.B. would like to thank Ugo Bruzzo and Barbara Fantechi for useful discussions and advices.
R.S. would like to thank for the same reason Andr\'es Collinucci and Roberto Valandro.
This research was supported in part by the Project of Knowledge Innovation
Program (PKIP) of Chinese Academy of Sciences, Grant No. KJCX2.YW.W10.

\end{document}